\RequirePackage{fix-cm}  %

\documentclass[twocolumn,epjc3]{svjour3}
\smartqed
\RequirePackage{graphicx}
\RequirePackage{mathptmx}
\RequirePackage{amsmath}
\RequirePackage{amsfonts}
\RequirePackage{amssymb}
\RequirePackage{latexsym}
\RequirePackage{widetext}

\journalname{Eur. Phys. J. C}

\begin{document}
\title{Homogeneous G\"{o}del-type solutions in hybrid metric-Palatini gravity}

\author{J. Santos\thanksref{e1,addr1}
        \and
        M. J. Rebou\c{c}as\thanksref{e2,addr2}
        \and
        A.F.F. Teixeira\thanksref{addr2}  }

\thankstext{e1}{e-mail: janilo.santos@gmail.com}
\thankstext{e2}{e-mail: reboucas@cbpf.br}

\institute{Departamento de F\'{\i}sica Te\'orica e Experimental,
Universidade Federal do Rio Grande do Norte, 59072-970 Natal -- RN, Brazil \label{addr1}
           \and
Centro Brasileiro de Pesquisas F\'{\i}sicas, Rua Dr.\ Xavier Sigaud 150,
22290-180 Rio de Janeiro -- RJ, Brazil \label{addr2}   }

\date{Received: date / Accepted: date}

\maketitle
\sloppy

\begin{abstract}
The hybrid metric-Palatini $f(\mathcal{R})$ gravity is a recently devised approach
to modified gravity in which it is added to the metric Ricci scalar $R$, in the
Einstein-Hilbert Lagrangian, a function $f(\mathcal{R})$ of Palatini curvature
scalar $\mathcal{R}$, which is constructed from an independent connection. %
These hybrid metric-Palatini gravity theories provide an alternative way to explain
the current accelerating expansion without a dark energy matter component.
If gravitation is to be described by a hybrid metric-Palatini $f(\mathcal{R})$ gravity
theory there are a number of issues that ought to be examined in its context,
including the question as to whether its equations allow homogeneous G\"odel-type
solutions, which necessarily leads to violation of causality.
Here, to look further into the potentialities and difficulties of $f(\mathcal{R})$
theories, we examine whether they admit  G\"odel-type solutions for physically
well-motivated matter source.
We first show that under certain conditions on the matter sources the problem
of finding out space-time homogeneous (ST-homogeneous) solutions in
$f(\mathcal{R})$ theories reduces to the problem of determining solutions
of Einstein's field equations with a cosmological constant.
Employing this far-reaching result, we determine a general ST-homo\-ge\-neous
G\"odel-type solution whose matter source is a combination of a scalar with an
electromagnetic fields plus a perfect fluid.
This general G\"odel-type solution contains special solutions in which the
essential parameter $m^2$ can be $m^{2} > 0$ hyperbolic family, $m=0$
linear class, and $m^{2} < 0$ trigonometric family, covering thus all
classes of homogeneous G\"odel-type spacetimes.
This general solution also contains all previously known solutions as
special cases.
The bare existence of these G\"odel-type solutions makes
apparent that hybrid metric-Palatini $f(\mathcal{R})$ gravity does not
remedy causal anomaly in the form of closed timelike curves that
are permitted in general relativity.

\end{abstract}

\section{Introduction}

A number of cosmological observations coming from different sources,
including the supernovae type Ia (SNe Ia)~\cite{SNE,SNE-1,SNE-2}, the cosmic
microwave background radiation (CMBR)~\cite{CMB,CMB-1} and  baryon acoustic
oscillation (BAO) surveys~\cite{BAO,BAO-1,BAO-2,BAO-4,BAO-5},
indicate that the Universe is presently expanding with an accelerating rate.
The frameworks proposed to account for this observed accelerated expansion can be
roughly grouped into two families. In the first, the underlying theory, general relativity
(GR), is kept unchanged, and the so-called dark energy component is invoked.
In this context, the simplest way to account for the  accelerating expansion of
the Universe is through the introduction of a cosmological constant, $\Lambda$, into
Einstein's field equations. This is completely consistent with the available observational
data, but it faces difficulties such as the order of magnitude of the cosmological constant
and its microphysical origin.
In the second family, modifications of Einstein's field equations are assumed
as an alternative for explaining the accelerated expansion. This latter group
includes, for example, generalized theories of gravity based upon modifications of
the Einstein-Hilbert action by taking nonlinear functions, $f(R)$, of the Ricci
scalar $R$  or other curvature invariants (for reviews see Refs.~\cite{fr-reviews,fr-reviews-1,fr-reviews-2,fr-reviews-3,fr-reviews-4,fr-reviews-5}).

In dealing with $f(R)$ gravity theories two different variational approaches,
which give rise to different dynamics, have often been considered in the literature%
~\cite{fr-reviews,fr-reviews-1,fr-reviews-2,fr-reviews-3,fr-reviews-4,fr-reviews-5}.
In the so-called metric formalism the connection is assumed to be Levi-Civita, therefore
defined by the metric.
In the Palatini formalism the metric and the connection are treated as independent fields,
and it is assumed that the matter fields do not couple with the independent connections.
Although these approaches have been invoked as possible ways to satisfactorily deal with
the observed accelerated expansion of the Universe, it has been pointed out that $f(R)$
gravity theories can face relevant difficulties,  including  the evolution of
cosmological perturbations and local gravity constraints (see, for example, Refs.~\cite{Koivisto-1,Hu,Tsujikawa,Koivisto-2,fr-reviews-3,Olmo-1,Olmo-2}).

These undesirable features have motivated a recent approach to modified $f(R)$
theories of gravity, which can be employed as a possible way to handle the
observed late-time cosmic acceleration, and also circumvent some difficulties
that arise in the framework of $f(R)$ theories in both formalisms.
The hybrid metric-Palatini $f(\mathcal{R})$ gravity is a recently devised approach to
such modified theories, in which it is added to the ordinary Ricci scalar $R$, in the
Einstein-Hilbert Lagrangian, a function, $f(\mathcal{R})$, of Palatini curvature
scalar $\mathcal{R}$, which is constructed from the independent connection
$\Gamma_{\mu\nu}^{\rho}$~\cite{Harko}.
These hybrid metric-Palatini gravity theories appears to suitably unify the description of the
late-time cosmic acceleration with the local solar system constraints~\cite{Harko,capozzi-Harko}.
Some of the astrophysical and cosmological implications of hybrid metric-Palatini gravity have
been examined in a number of papers~\cite{capozzi-1,capozzi-2,capozzi-3,capozzi-4,Lima-1}.
Wormwhole solutions, Einstein static universe, linear perturbations, the Cauchy problem,
dynamical system analysis and a brane model have been discussed, respectively, in the references%
~\cite{capozzi-5,Bohmer,Lima-2,capozzi-6,Carloni,Qi-Ming}. Other important matters such as
Noether symmetries~\cite{Borowiec} and the thermodynamic behavior~\cite{Azizi} have also
been recently considered. For an introduction to hybrid metric-Palatini $f(\mathcal{R})$
gravity and a detailed list of related references, we refer the reader to the recent
review article~\cite{capozzi-4}.

In general relativity (GR) the space-times have \textit{locally} the same causal
structure of the flat space-time of special relativity since
the space-times of GR are locally Minkowskian.
On nonlocal scale, however, significant differences may arise since
the general relativity field equations do no provide
nonlocal constraints on the underlying space-times.
Indeed, it has long been known that there are solutions to Einstein's
field equations that present nonlocal causal anomalies in the form of closed
time-like curves (see, for example,
Refs.~\cite{CTLC-examples,CTLC-examples-1,CTLC-examples-2,%
CTLC-examples-3,CTLC-examples-4,CTLC-examples-5}.  

The renowned model found by G\"odel~\cite{Godel49} is the best known example of
a solution to the Einstein's equations, with a physically well-motivated
source, that makes it apparent that GR permits solutions with closed timelike
world lines, regardless of its local Lorentzian character that ensures locally
an inherited regular chronology and therefore the local validation of the
causality principle.
The G\"odel model is a solution of Einstein's equations with cosmological
constant $\Lambda$ for dust of density $\rho$, but it can also be viewed as
perfect-fluid solution with equation of state $p=\rho\,$ with no
cosmological constant term.
Owing to its unexpected properties, G\"odel's solution has a
recognizable importance and has motivated a considerable  
number of investigations on rotating G\"odel-type models
as well as on causal anomalies in the context of general 
relativity (see, e.g.
Refs.~\cite{CTLC-examples,CTLC-examples-1,CTLC-examples-2,%
CTLC-examples-3,CTLC-examples-4,CTLC-examples-5,Godel49,%
GT_in_GR,GT_in_GR-1,GT_in_GR-2,GT_in_GR-3,GT_in_GR-4,%
GT_in_GR-5,GT_in_GR-7,GT_in_GR-8,GT_in_GR-10,GT_in_GR-11}
and  other gravity theories~\cite{GT,GT-1,GT-2,GT-3,GT-4,%
GT-5,GT-6,GT-7,GT-8,GT-9,GT-10,GT-11,GT-12,GT-13,GT-14,GT-15,RS-2009,RST-2010,R-47,%
R-49,FonPetrovReb,R-55,GT-16,GT-17,Otalora-Reboucas-2017}.
In two recent papers, we have also examined G\"odel-type models and the violation
of causality problem for $f(R)$ gravity in both the metric and Palatini
variational approaches~\cite{RS-2009,RST-2010}, extending  therefore
the results of Refs.~\cite{CliftonBarrow2005} and~\cite{Reb_Tiomno}.

If gravitation is to be described by hybrid metric-Palatini $f(\mathcal{R})$
gravity theory there are a number of issues that ought to be reexamined in its
context, including its consistence with the recent detection of gravitational
wave~\cite{GW-detec} and the question as to whether these gravity theories allow
G\"odel-type solutions, which necessarily lead to closed timelke curves, or would
remedy this causal pathology by ruling out this type of solutions, which are permitted
in general relativity.

In this article, to proceed further with the investigations on the potentialities,
difficulties and limitations of $f(\mathcal{R})$,  we undertake this question
by examining whether the $f(\mathcal{R})$ gravity theories admit homogeneous
G\"odel-type solutions for a combination of physically well-motivated
matter sources.
To this end, we first examine the general problem of finding out ST-homogeneous solutions
in hybrid metric-Palatini $f(\mathcal{R})$ gravity for matter sources with constant
trace $T$ (scalar) of the energy-momentum tensor, and show that it reduces to
to the problem of determining ST-homogeneous solutions
of Einstein's field equations with a cosmological constant determined by
$f(\mathcal{R})$ and its first derivative $f'(\mathcal{R})$.
Employing this far-reaching result, we determine a general ST-homogeneous G\"odel-type
whose matter source is a combination of a scalar with an electromagnetic fields
plus a perfect fluid.
This general G\"odel-type solution contains special solutions in which the
essential parameter $m^2$ defines any one of the possible classified families
homogeneous G\"odel-type solutions, namely $m^{2} > 0$ hyperbolic family, $m=0$
linear class, and $m^{2} < 0$ trigonometric family.
This general homogeneous G\"odel-type solution also contains previously known
solutions as special cases.
There emerges from one of the particular solution of the hyperbolic family
that every perfect-fluid G\"{o}del-type solution of any  $f(\mathcal{R})$
gravity with density $\rho$ and pressure $p$ and satisfying the weak energy
conditions $\rho >0 \; \mbox{and}\; \rho+p \geq 0$ is necessarily isometric
to the G\"odel geometry%
\footnote{
This extends to the context of  $f(\mathcal{R})$  gravity a
theorem which states that every perfect-fluid G\"odel-type solution
of Einstein's equations is necessarily isometric to the G\"odel
spacetime~\cite{BampiZordan78}.}.

The bare existence of these noncausal G\"odel-type solutions makes
apparent that hybrid metric-Palatini $f(\mathcal{R})$ gravity does not
remedy causal anomaly in the form of closed timelike curves that
are permitted in general relativity.

The structure of the paper is as follows.  In Section~\ref{Hyb-grav-th}
we give a brief account of the hybrid metric-Palatini  $f({\mathcal{R}})$
gravity theories.
In Section~\ref{GT-geo} we present the basic properties of homogenous
G\"{o}del-type geometries. This includes the conditions for space-time
homogeneity, a classification of ST-homogeneous G\"{o}del-type geometries
and a study of the existence of closed time-like curves in all ST-homogeneous
G\"odel-type metrics.
In Section~\ref{Homo-sol} we first examine the problem of finding out
ST-homogeneous solutions in  $f(\mathcal{R})$ gravity whose trace $T$
of the energy-momentum tensor of the matter source  is constant, and
show that in such cases the problem reduces to that of finding out solutions
of Einstein's field equations with a cosmological constant.
We then show that the hybrid metric-Palatini $f(\mathcal{R})$  gravity
theories admit ST-homogeneous G\"{o}del-type solutions for a
general combination of physically well-motivated matter contents.

\section{Hybrid metric-Palatini gravity} \label{Hyb-grav-th}

The action that defines a hybrid metric-Palatini gravity is given by
\begin{equation}
\label{actionJF}
S =\dfrac{1}{2\kappa^{2}}\int d^4x\sqrt{-g}\,\left[ R + f({\mathcal{R}}) + \mathcal{L}_{m}\right]\,,
\end{equation}
where  $\kappa^2=8\pi G$, $g$ is the determinant of the metric tensor $g_{\mu \nu}\,$,
$R$ is the Ricci scalar associated to the Levi-Civita connection of the metric $g_{\mu\nu}$,
$\mathcal{L}_{m}$ is the Lagragian density for the matter fields, and the extra term
$f({\mathcal{R}})$ is a function of Palatini curvature scalar ${\mathcal{R}}$, which depends
on the metric and on an independent connection  $\Gamma_{\mu\nu}^{\rho}$ through
\begin{equation} \label{connect}
{\mathcal{R}}\equiv g^{\mu\nu}\mathcal{R}_{\mu\nu} =
g^{\mu\nu}\left( \partial_{\rho}\Gamma_{\mu\nu}^{\rho} - \partial_{\nu}\Gamma_{\mu\rho}^{\rho} +
\Gamma_{\rho\lambda}^{\rho}\Gamma_{\mu\nu}^{\lambda} -
\Gamma_{\mu\lambda}^{\rho}\Gamma_{\rho\nu}^{\lambda}\right).
\end{equation}
The variation of the action~(\ref{actionJF}) with respect to the metric gives the
field equations
\begin{equation} \label{field_eq}
G_{\mu\nu} + F(\mathcal{R})\mathcal{R}_{\mu\nu} - \frac{f(\mathcal{R})}{2}g_{\mu\nu}
 = \kappa^2T_{\mu\nu}\,,
\end{equation}
where $G_{\mu\nu}=R_{\mu\nu} - R/2\,g_{\mu\nu}$ and  $R_{\mu\nu}$ are, respectively,
Einstein and  Ricci tensor associated with the Levi-Civita connection of $g_{\mu\nu}$,
$F(\mathcal{R})\equiv df/d{\mathcal{R}}$,
$T_{\mu\nu}=-2/\sqrt{-g}\;\,\delta (\sqrt{-g}\mathcal{L}_m )/\delta g^{\mu\nu}$ is the
energy-momentum tensor of the matter fields.

The variation of the action~(\ref{actionJF}) with respect to the independent connection
$\Gamma_{\mu\nu}^{\rho}$ yields
\begin{equation}  \label{connections_eq}
\widetilde{\nabla}_\beta\left( \sqrt{-g}\,F(\mathcal{R})\,g^{\mu\nu} \right) = 0 \,,
\end{equation}
where  $\widetilde{\nabla}_\beta$ denotes the covariant derivative associated with
$\Gamma_{\mu\nu}^{\rho}$.
If one defines a metric $h_{\mu\nu} = F(\mathcal{R})\,g_{\mu\nu}$, it can be easily
shown that Eq.~(\ref{connections_eq}) determines a Levi-Civita connection of $h_{\mu\nu}$,
which in turn can be rewritten in terms of $g_{\mu\nu}$ and its Levi-Civita connection
 $\left\{^{\rho}_{\mu\nu}\right\}$ in the form
\begin{equation}  \label{Gamma}
\Gamma_{\mu\nu}^{\rho} = \left\{^{\rho}_{\mu\nu}\right\}
+ \frac{1}{2}\left( \delta^{\rho}_{\mu}\partial_{\nu}
+ \delta^{\rho}_{\nu}\partial_{\mu}
- g_{\mu\nu}g^{\rho\sigma}\partial_{\sigma} \right)\ln F(\mathcal{R})\,.
\end{equation}
Using now Eq.~(\ref{Gamma}) one finds the relation between the two Ricci tensors, which is
given by
\begin{equation} \label{Ric-relation}
\mathcal{R}_{\mu\nu} = R_{\mu\nu} + \frac{3}{2F^2}\,\partial_{\mu}\,F\,\partial_{\nu}\,F
- \frac{1}{F}(\nabla_{\mu}\nabla_{\nu} + \frac{1}{2}g_{\mu\nu}\Box)F\,,
\end{equation}
where $\nabla_{\mu}$ denotes the covariant derivative associated to
$\left\{^{\rho}_{\mu\nu}\right\}$.

Equation~(\ref{Ric-relation}) in turn gives rise to the following relation between
the two Ricci scalars:
\begin{equation} \label{scalar-relation}
\mathcal{R} = R + \frac{3}{2F^2}(\partial F)^2 - \frac{3}{F}\Box F\,,
\end{equation}
where $(\partial F)^2 = g^{\alpha\beta}\,\partial_{\alpha}F\partial_{\beta}F$ and
$\Box F = g^{\alpha \beta}\,\nabla_{\alpha}\nabla_{\beta}\,F$.

The Palatini curvature $\mathcal{R}$ can be obtained from the trace of the
field Eq.~(\ref{field_eq}), which yields
\begin{equation} \label{trace}
\mathcal{R}F(\mathcal{R}) - 2f(\mathcal{R}) = \kappa^2T + R \equiv X\,.
\end{equation}
This trace equation can be used to express $\mathcal{R}$ algebraically in terms
of $X$ when the $f(\mathcal{R})$ is given as an analytic expression.
Finally, we note that the variable $X$ measures the deviation from the
general relativity trace equation $R=-\kappa^2T$.

\section{Homogeneous G\"{o}del-type geometries} \label{GT-geo}

To make this work clear and to a certain extent self-contained, in this section we
present the basic properties of homogenous G\"{o}del-type geometries, which
we use in the following sections. To this end, we first discuss the conditions
for space-time homogeneity (ST-homogeneity) of these space-times, and present
all non-isometric ST-homogeneous G\"odel-type classes. These ST-homogeneity conditions
along with the set of isometrically non-equivalent geometries are important in
the determination of ST-homogeneous G\"odel-type solutions, in Section~\ref{Homo-sol},
to the hybrid metric-Palatini $f(\mathcal{R})$ field equations.
Second, we discuss the existence of closed time-like curves in the metrics of
these classes. The existence of these non-causal curves are crucial to examine
whether hybrid metric-Palatini $f(\mathcal{R})$ 
gravities allow violation of causality of G\"odel-type.

\subsection{Homogeneity and non-equivalent metrics}

G\"odel solution to the general relativity field equations is a particular
member of the broad family of  geometries whose general form in cylindrical
coordinates, $(r, \phi, z)$, is given by~\cite{Reb_Tiomno}
\begin{equation}  \label{G-type_metric}
ds^2 = [dt + H(r)d\phi]^2 - D^2(r)d\phi^2 - dr^2 - dz^2\,.
\end{equation}
The necessary and sufficient conditions for the G\"odel-type metric~(\ref{G-type_metric})
to be space-time homogeneous (ST-homogeneous) are given by~\cite{Reb_Tiomno,RebAman}
\begin{equation} \label{ST-hom-cond}
\frac{H'}{D}  =  2\omega \qquad \; \text{and} \qquad \;
\frac{D''}{D}  =  m^{2},
\end{equation}
where the prime denote derivative with respect $r$, and the parameters $(\Omega,m)$
are constants such that  $\Omega^{2}>0$ and $-\infty\leq m^{2}\leq\infty$.

As a matter of fact, except for the case $m^2= 4 \omega^2$ all locally ST-homogeneous
G\"odel-type space-times admit a group  $G_{5}$  of isometries acting transitively on
the whole space-time~\cite{RebAman}. The special case $m^2= 4 \omega^2$ admits a
$G_{7}$ of isometries~\cite{TeiRebAman,RebAman}.

The irreducible set of isometrically nonequivalent ST-homogeneous G\"odel-type
metrics can be obtained by integrating equations~(\ref{ST-hom-cond})
and suitably eliminating nonessential integration constants. The final result is
that ST-homogeneous G\"odel-type geometries can be grouped in the following three
classes~\cite{Reb_Tiomno}:
\begin{enumerate}
\item[\bf i.] Hyperbolic, in which $m^{2} = \mbox{const} > 0$ and
\begin{equation} \label{HD-hyperb}
H =\frac{4\,\omega}{m^{2}} \,\sinh^2 (\frac{mr}{2}), \;\;\;\;
               D=\frac{1}{m} \sinh\,(mr)\,;
\end{equation}
\item[\bf ii.] Linear, in which $m=0$ and
\begin{equation} \label{HD-linear}
 H = \omega r^{2},  \;\;\;\; D=r  \,,
\end{equation}
\item[\bf iii.] Trigonometric,  where $m^{2}= \mbox{const} \equiv - \mu^{2} < 0 $\  and
\begin{equation} \label{HD-circul}
 H = \frac{4\,\omega}{\mu^{2}} \,\sin^2 (\frac{\mu r}{2}),
                            \;\;\;\; D=\frac{1}{\mu} \sin\,(\mu r)\,.
\end{equation}
\end{enumerate}
Thus, clearly all ST-homogeneous G\"odel-type geometries are characterized by the
two independent parameters $m^2$ and  $\omega$ --- identical pairs $(m^2, \omega^2)$
specify isometric spacetimes~\cite{Reb_Tiomno,TeiRebAman,RebAman}.%
\footnote{G\"odel geometry is a solution of Eintein's equations in general relativity,
is indeed a particular case of the hyperbolic class of geometries in which $m^2= 2 \omega^2$.}
In this way, to determine whether hybrid metric-Palatini $f(\mathcal{R})$ gravity
allows G\"odel-type solutions is to find out whether its field equations can be used
to specify a pair of parameters $m^2$ and  $\omega$ for a suitably chosen matter
source.

\subsection{Closed time-like curves}  \label{CTLC}

We begin by noting that the presence of a single closed timelike curve
in a space-time is an unequivocal manifestation of violation of causality.
However, a space-time may admit non-causal closed curves other than
G\"odel's circles we discuss in this Section.
To examine the existence of closed time-like curves in ST-homogeneous
G\"odel-type metrics we first rewrite the line element~(\ref{G-type_metric})
as
\begin{equation} \label{G-type_metric2}
ds^2=dt^2 +2\,H(r)\, dt\,d\phi -dr^2 -G(r)\,d\phi^2 -dz^2 \,,
\end{equation}
where $G(r)= D^2 - H^2$.
In this form it is easy to show that existence of closed time-like curves,
which allows for violation of causality in homogeneous G\"odel-type space-times,
depends on the sign of  the metric function $G(r)$.
Indeed, from Eq.~(\ref{G-type_metric2}) one has that the circles, hereafter
called G\"odel's circles,  defined
by $t, z, r = \text{const}$ become closed timelike curves whenever $G(r) < 0$.

For the hyperbolic ($m^2>0$) class of homogeneous G\"odel-type metrics, from
Eqs.~(\ref{HD-hyperb}) one has that
\begin{equation}
G(r) = \frac{4}{m^2} \, \sinh^2 (\frac{mr}{2}) \left[ ( 1- \frac{4\omega^2}{m^2})\,
\sinh^2 (\frac{mr}{2})+1 \right]\,.
\end{equation}
Therefore for $0 < m^2 < 4\omega^2$ there is a critical radius $r_c$ defined
by $G(r)=0$, which is given by
\begin{equation} \label{r-critical}
\sinh^2 \frac{mr_c}{2}= \left[ \frac{4\omega^2}{m^2} - 1 \right]^{-1}\,,
\end{equation}
such that for $r<r_{c}$ one has $G(r)>0$, and  for $r>r_{c}$ one has $G(r)<0$.
Thus, the circles $t,r,z=\text{const}$   with  $r>r_{c}$ are closed timelike curves.%
\footnote{The only ST-homogeneous G\"odel-type space-time without these
noncausal circles come about when $m^2 = 4 \omega^2$ (see Ref.~\cite{Reb_Tiomno}).
In this case, the critical radius $r_c \rightarrow \infty$,
and hence the violation of causality of G\"odel type is avoided.}

For linear class ($m=0$) of homogeneous G\"odel-type space-times,
from Eq.~(\ref{HD-linear}) one easily finds
\begin{equation}
G(r)=r^{2}-r^{4}\,\omega^{2}=-r^{2}\,\left(r\,\omega-1\right)\,
\left(r\,\omega+1\right) \,.\label{eq:G linear}
\end{equation}
Thus, there is a critical radius, defined by $\,G(r)=0\,$, and
given  by $r_c = 1/\omega$, such that for any radius $r > r_c$ one
has $G(r)<0$, and then the circles defined by $t, z, r = \text{const}$
are closed timelike curves.

Finally for the trigonometric class ($m^{2}= \mbox{const} \equiv - \mu^{2} < 0 $),
from the metric functions given by Eq.~(\ref{HD-circul}) one finds
\begin{equation}
G\left(r\right)=\frac{4\,}{\mu^{4}}\sin^{2}(\frac{\mu\, r}{2})\,[\,\mu{}^{2}
-(4\,\omega^{2}+\mu{}^{2})\, \sin^{2}(\frac{\mu\, r}{2})\,] \,,\label{eq:G trig}
\end{equation}
and therefore  $G(r)$  has an infinite sequence of zeros. Thus, in the section
$t, z, r = \text{const}$,  there is an sequence of alternating causal [$\,G(r) > 0\,$]
and noncausal [$\,G(r) < 0\,$] regions without and with noncausal circles, depending on
the value of $r=\text{const}$ (For more details see the Appendix
of Ref.~\cite{FonPetrovReb}).
In this way, if $G(r) < 0$ for a certain range of $r$ ($ r_1 < r < r_2$, say)
noncausal G\"odel's circles exist, whereas for $r$ in the next circular band
$r_2 < r < r_3$ (say) for which $G(r) > 0$ no such closed timelike circles
exist, and so on.

To close this section, we note that in this paper by non-causal and
causal solutions we mean, respectively, solutions with and
without violation of causality of G\"odel-type, i.e., with
and without G\"odel's circles.

\section{Solutions in hybrid metric-Palatini  gravity} \label{Homo-sol}

The aim of this section is twofold. First, we examine the problem of finding out
ST-homogeneous solutions in hybrid metric-Palatini $f(\mathcal{R})$ gravity
whose trace $T$ of the energy-momentum tensor of the matter source  
is constant. We show that in such cases the problem of finding out solutions in the hybrid
metric-Palatini $f(\mathcal{R})$ gravity reduces to the problem of determining
ST-homogeneous solutions of Einstein's field equations with a cosmological constant
determined by $f(\mathcal{R})$ and its first derivative $F = f'(\mathcal{R})$.
Second, we examine whether hybrid metric-Palatini $f(\mathcal{R})$  field equations
admit ST-homogeneous G\"{o}del-type solutions for a combination of a scalar field with
an electromagnetic field plus a perfect fluid.

\subsection{Field equations} \label{Field-Eqs-Hom}

We begin by noting that using equation~(\ref{Ric-relation}) the field
equations~(\ref{field_eq})  of the hybrid metric-Palatini $f(\mathcal{R})$
gravity can be rewritten in the form
\begin{equation} \label{eff-field-eqs}
G_{\mu\nu} = \kappa^2\,T^{\rm{eff}}_{\mu\nu} =
\kappa^2\left(\, T_{\mu\nu} + T_{\mu\nu}^{\mathcal{R}}\, \right)\,,
\end{equation}
where
\begin{eqnarray} \label{energy-eff}   
\kappa^2 T_{\mu\nu}^{\mathcal{R}} = \frac{1}{2}\left[ f(\mathcal{R}) + \Box F(\mathcal{R}) \right]\,g_{\mu\nu} - F(\mathcal{R})R_{\mu\nu} \nonumber \\
+ \nabla_{\mu}\nabla_{\nu}F(\mathcal{R})
-\frac{3}{2F(\mathcal{R})}\,\partial_{\mu}F(\mathcal{R})\partial_{\nu}F(\mathcal{R})\,.
\end{eqnarray}
In this context, an important constraint comes from the trace of the field
equations~(\ref{eff-field-eqs}), which can be written in the form
\begin{equation}
R + \kappa^2T = X = - \kappa^2T^{\mathcal{R}}\,,
\end{equation}
where from equation~(\ref{energy-eff}) one has
\begin{equation}\label{eff-trace}
\kappa^2T^{\mathcal{R}} = 2f(\mathcal{R}) - RF(\mathcal{R}) + 3\Box F(\mathcal{R}) - \frac{3}{2}\,\frac{[\partial F(\mathcal{R})]^2}{F(\mathcal{R})}\,.
\end{equation}

For ST-homogeneous spacetimes, which we are concerned with in this paper, one has
that the Ricci scalar is necessarily constant. On the other hand, for matter sources
whose trace of  the energy-momentum  $T$ is also constant, which we focus in this paper,
one has $X = \mbox{const}= - \kappa^2T^{\mathcal{R}}$.
In such cases, one can show that the field equations of the hybrid
metric-Palatini gravity reduces to Einstein's field equations with a cosmological
constant. 

To this end, we first examine the second term on the right hand side of
equation~(\ref{Gamma}) which gives the departures of the independent connection 
$\Gamma_{\mu\nu}^{\rho}$ from Levi-Civita connection $\left\{^{\rho}_{\mu\nu}\right\}$.
Clearly,  each individual part of this second term in this equation is proportional to
\begin{equation} \label{duln}
\partial_{\mu}\ln{F(\mathcal{R})} = \frac{F'}{F} \,\partial_{\mu} \mathcal{R}
=\frac{1}{F} \,\,\partial_{\mu} F(\mathcal{R}) \,.
\end{equation}
On the other hand, to calculate $\partial_{\mu} F(\mathcal{R})$ we note that from the
trace equation~(\ref{trace}) one has
$\partial_{\mu} \mathcal{R} = \partial_{\mu} X \, / \,[\mathcal{R} F'(\mathcal{R})
- F(\mathcal{R})]$,
which together with equation~(\ref{duln}) furnishes 
\begin{equation}\label{partial-derivatives}
\partial_{\mu}F(\mathcal{R}) = \frac{F'(\mathcal{R})\,\partial_{\mu}X}{\mathcal{R}F'(\mathcal{R})
 - F(\mathcal{R}) } \,,  
\end{equation}
provided that $\mathcal{R}F'(\mathcal{R})-F(\mathcal{R})\neq 0$.%
%
%
{}From equations~(\ref{duln}) and~(\ref{partial-derivatives}) one has that for $X = \mbox{const}$
the connection $\Gamma_{\mu\nu}^{\rho}$ reduces to Levi-Civita connection
$\left\{^{\rho}_{\mu\nu}\right\}$. Furthemore,  from equations~(\ref{Ric-relation})
and~(\ref{scalar-relation}) one can easily show that $\partial_{\mu}F(\mathcal{R})= 0$ also
ensures that $\mathcal{R}_{\mu\nu}=R_{\mu\nu}$ and $\mathcal{R}=R$.
This makes apparent that ST-homogeneous spacetimes solutions whose trace of the energy-momentum
tensor  is constant ($R + \kappa^2T = X = {\rm const} $) the field equations of the hybrid
metric-Palatini gravity~(\ref{field_eq}) reduce formally to field equations of  $f(R)$ theories
in the metric formalism, which can clearly be rewritten in the form
\begin{equation} \label{fiel_eq-X_II}
\left[1 + F(R)\right]\,G_{\mu\nu}  - \frac{1}{2}\,\left[\,f(R)- RF(R)\,\right]\,g_{\mu\nu}
=  \kappa^2\,T_{\mu\nu}\,,
\end{equation}
with associated trace equation 
\begin{equation}\label{constraint}
\qquad R F(R) - 2f(R) = \kappa^2T + R = \mbox{const.}
\end{equation}

However, for an explicitly given $f(\mathcal{R})$,  solving the algebraic
equation~\eqref{constraint}  one finds constant roots $R\,$'s.
Thus, for each explicit root the field equations \eqref{fiel_eq-X_II} can be
rewritten in the form%
\footnote{Clearly, different roots $R$ give rise to different
rescales of $\kappa^2$, and different effective cosmological
constant $\Lambda$.}

\begin{equation} \label{Einstein-Lambda}
G_{\mu\nu} =  \bar{\kappa}^2 \, T_{\mu\nu} + \Lambda \, g_{\mu\nu}  \,,
\end{equation}
where
\begin{equation}  \label{Lambda-kappa}
\Lambda = \frac{f(R) - R\, F(R)}{2\,[1+F(R)]} \quad \mbox{and} \quad
\bar{\kappa}^2 = \frac{\kappa^2}{1+F(R)}   \,\,.
\end{equation}
The trace equation becomes
\begin{equation}  \label{constraint-2}
\qquad \quad  R + \bar{\kappa}^2\,T + 4 \Lambda = 0 \,.
\end{equation}
Clearly, the factor $[1+F(R)]$ in equations \eqref{Lambda-kappa} is a constant
that simply rescales the units of $\kappa^2$ and the effective cosmological
constant $\Lambda$.

\subsection{G\"odel-type solutions} \label{sols} 

In this section we discuss ST-homogeneous G\"odel-type solutions in hybrid metric-Palatini
$f(\mathcal{R})$ gravity for well-motivated matter contents whose trace of the
energy-momentum tensor is constant.

We begin by noting that the search for ST-homogeneous G\"odel-type solutions to the
hybrid metric-Palatini gravity field equations is greatly simplified if instead of using
coordinates basis one uses a new basis given by the following set of linearly
independent one-forms (tetrad frame) $\Theta^A$:
\begin{equation} \label{one_forms1}
\theta^0 = dt + H(r)d\phi\,, \;  \theta^1 = dr\,,  \;
\theta^2 = D(r)d\phi\,, \;  \theta^3 = dz  \,,
\end{equation}
relative to which the G\"odel-type line element~(\ref{G-type_metric}) takes
the local Lorentzian form
\begin{equation}  \label{G-type_metric3}
ds^2 = \eta_{AB}\,\theta^A\theta^B =
(\theta^0)^2 - (\theta^1)^2 - (\theta^2)^2 - (\theta^3)^2\,.
\end{equation}
Here and in what follows capital letters are tetrad indices (or Lorentz frame indices)
and run from 0 to 3. These Lorentz frame indices are raised and lowered with Lorentz
matrices $\eta^{AB} = \eta_{AB} = \mbox{diag} (+1, -1, -1, -1)$, respectively.

In the tetrad frame~(\ref{one_forms1}) the nonvanishing components of
the Ricci tensor, $R_{AB}= \eta^{CD} R_{CADB}$, are given by
\begin{eqnarray}
R_{02} &=&  \frac{1}{2} \, \left( \frac{H'}{D}\, \right)' \,, \quad
R_{00} = \frac{1}{2} \, \left( \frac{H'}{D}\, \right)^2 \,, \label{Ric-02-00} \\
R_{11} &= & R_{22} = \frac{1}{2} \, \left( \frac{H'}{D}\, \right)^2
- \frac{D''}{D}  \label{Ric-11-22} \,,
\end{eqnarray}
where the prime denotes derivative with respect to $r$.
Since the Lorentz frame components of the Ricci tensor depend only on $H'/D$ and $D''/D$,
from the ST-homogeneity conditions~(\ref{ST-hom-cond}) one has that for all classes
of ST-homogeneous G\"odel-type metrics the frame components of the Ricci tensor are constants.
Thus, the Ricci scalar is also constant and given by  $R = \eta^{CD} R_{CD}= 2  (m^2 - \omega^2)$.

In this Lorentzian basis the field equations~(\ref{Einstein-Lambda}) reduce to
\begin{equation}  \label{G_AB}
G_{AB} =  \bar{\kappa}^2 \, T_{AB} + \Lambda \,\eta_{AB}\,,
\end{equation}
where from equations~(\ref{Ric-02-00}) and~(\ref{Ric-11-22}) along with the
conditions~(\ref{ST-hom-cond}) one has that the only nonvanishing Lorentz frame components of
the Einstein tensor $G_{AB}$ for ST-homogeneous G\"odel-type metrics take the very
simple form
\begin{equation} \label{GAB_components}
G_{00} =  3 \omega^2 - m^2\,, \quad
G_{11} = G_{22}  =  \omega^2\,, \quad
G_{33}  =  m^2 - \omega^2\,.
\end{equation}

\subsubsection{Combined-fields general solution }  \label{Comb-sol}

In this section we take combination of scalar and electromagnetic fields
with a perfect fluid as a matter source, and find a general ST-homogeneous
G\"odel-type solution, which contains the a perfect fluid and a scalar
field particular solutions, and whose essential parameter $m^2$ can
be $m^{2} > 0$ (hyperbolic family), $m=0$ (linear class) or $m^{2} < 0$
(trigonometric family) depending on the amplitude values of the matter
components.

In the Lorentzian basis~(\ref{one_forms1}) the energy-momentum tensor of
combined matter sources takes the form
\begin{equation}  \label{T-AB-tot}
T_{AB} =T^{(M)}_{AB} + T^{(S)}_{AB} + T^{(EM)}_{AB}  \,,
\end{equation}
where $T^{(M)}_{AB}$, $T^{(S)}_{AB}$ and $T^{(EM)}_{AB}$ are, respectively,
the energy momentum tensors of a perfect fluid, a scalar field, and
an electromagnetic field, which we discuss in what follows.

For a perfect fluid of density $\rho$ and pressure $p\,$,
$T^{(M)}_{AB}$ one has
\begin{equation} \label{perfect_fluid}
T^{(M)}_{AB} = (\rho + p)\,u_{A}u_{B} - p\,\eta_{AB}\,.
\end{equation}

The energy-momentum tensor of a single scalar field is given by
\begin{equation} \label{scalarfield}
T^{(S)}_{AB}= \Phi^{}_{|A}\,\Phi^{}_{|B} - \frac{1}{2}\,\eta^{}_{AB}\,
\Phi^{}_{|M} \,\Phi^{}_{|N}\, \eta^{MN} ,
\end{equation}
where vertical bar denotes components of covariant derivatives relative to
the local basis $\theta^A = e^{A}_{\ \alpha} \, dx^\alpha$ [see Eqs.~(\ref{one_forms1})
and (\ref{G-type_metric3})], i.e. $\Phi^{}_{|A} = e^A_{\ \mu} \,\nabla_{\mu} \Phi^{}$.
Following Ref.~\cite{Reb_Tiomno} it is straightforward to show that a scalar field
of the form $\Phi (z)= \varepsilon \,z + \epsilon$, with
$\varepsilon,\epsilon = \text{const}$, fulfills the scalar
field equation
$\Box \,\Phi = \eta^{AB}_{}\,\nabla_{A} \nabla_{B} \,\Phi\,=0$.
Thus,  the non-vanishing components of the energy-moment tensor for this scalar field
are
\begin{equation}  \label{S-comp}
T^{(S)}_{00} = - T^{(S)}_{11} = - T^{(S)}_{22} = T^{(S)}_{33} = \frac{\varepsilon^2}{2}\,,
\end{equation}

As for the electromagnetic part of energy momentum tensor \eqref{T-AB-tot},
following Ref.~\cite{Reb_Tiomno},  the electromagnetic field tensor $F_{AB}$
given by
\begin{eqnarray}\label{electromag-field}
&F_{03}=-F_{30} = E_0\sin[2\omega(z-z_0)]\,, \\
&F_{12}=-F_{21} = E_0\cos[2\omega(z-z_0)]\,,
\end{eqnarray}
satisfies the source-free Maxwell equations which, in the tetrad frame (\ref{one_forms1}),
take the form
\begin{eqnarray}\label{Maxwell-eq}
&F^{AB}\,_{|B} + \gamma^{A}\,\!_{MB}F^{MB} + \gamma^{C}\,\!_{MC}F^{AM} = 0\,, \\
&F_{[AB|C]} + 2F_{M[C}\gamma^{M}\,\!_{AB]} = 0\,,
\end{eqnarray}
where the brackets denote total anti-symmetrization and the Ricci rotation coefficients are
defined by $\gamma^{A}\,\!_{BC}=-\nabla^{}_\beta \theta^A_{\alpha} \,\,\theta^{\alpha}_B
\,\theta^{\beta}_C$.
The non-vanishing components of the associated energy-momentum tensor
$T^{(EM)}_{AB} = F_A^{\ C} \, F^{}_{BC} + \frac{1}{4} \eta^{}_{AB} F^{CD} F_{CD}$ are
given by
\begin{equation} \label{E-comp}
T^{(EM)}_{00} = T^{(EM)}_{11} = T^{(EM)}_{22} = -T^{(EM)}_{33} = \frac{E_0^2}{2}\,.
\end{equation}
Thus, taking into account equations~(\ref{GAB_components}), (\ref{perfect_fluid}),
(\ref{S-comp}) and (\ref{E-comp}), one has that for the combined-fields matter source
(\ref{T-AB-tot}) the field equations~(\ref{G_AB}) reduce to
\begin{eqnarray}
\qquad 3\omega^2 - m^{2} & = & \bar{\kappa}^2\left( \rho + \frac{\varepsilon^2}{2}
+ \frac{E^2_0}{2}\right) + \Lambda \,,   \label{1st-mix-Godel-field-eq} \\
\qquad \omega^{2} & = & \bar{\kappa}^2\left( p - \frac{\varepsilon^2}{2}
+ \frac{E^2_0}{2} \right) - \Lambda \,, \label{2nd-mix-Godel-field-eq} \\
\qquad m^2 - \omega^{2} & = & \bar{\kappa}^2 \left( p + \frac{\varepsilon^2}{2}
- \frac{E^2_0}{2} \right) - \Lambda \,.\label{3rd-mix-Godel-field-eq}
\end{eqnarray}

To determine the essential parameters $\omega^2$ and  $m^2$, we first
substitue~(\ref{3rd-mix-Godel-field-eq}) into~(\ref{1st-mix-Godel-field-eq})
to obtain
\begin{equation} \label{omega-mix-eq}
2\omega^2 =  \bar{\kappa}^2(\rho + p + \varepsilon^2) \,
\end{equation}
Second, we use~(\ref{2nd-mix-Godel-field-eq}), (\ref{3rd-mix-Godel-field-eq})
and~(\ref{omega-mix-eq}) to find
\begin{equation} \label{m-mix-eq}
m^2 = \bar{\kappa}^2 (\rho + p + 2 \varepsilon^2 - E_0^2) \,.
\end{equation}
The trace equation (\ref{constraint-2}) gives rise to the
following constrain for the cosmological constant
\begin{equation}\label{Lambda-constraint-eq}
2\,\Lambda = \bar{\kappa}^2 (p - \rho -2\varepsilon^2 + E_0^2 )
\end{equation}

Since ST-homogeneous G\"odel-type geometries are characterized by the
two essential parameters $m^2$ and  $\omega^2$, the above equations
(\ref{omega-mix-eq}) and (\ref{m-mix-eq})
make explicit how the  $f(\mathcal{R})$  gravity  specifies a pair
of parameters $(m^2, \omega^2)$, and therefore determines a general
ST-homogeneous G\"odel-type solution, for the combined-fields matter
source~(\ref{T-AB-tot}).

The general solution given by equations~(\ref{omega-mix-eq}) and~(\ref{m-mix-eq})
contains all general relativity G\"odel-type known  solutions~\cite{Reb_Tiomno}
as particular cases. Indeed, the perfect fluid G\"odel solution is
recovered when $\varepsilon = E_0 = 0$ with $\rho,p \neq 0 $,
whereas the scalar field causal solution~~\cite{Reb_Tiomno} is retrieved for
$\rho =p=E_0=0$ with $\varepsilon\neq 0$.%
\footnote{
We note that G\"{o}del metric ($m^2 = 2\omega^2$) can also be generated through
a particular combination of perfect fluid with a scalar field, namely for non-vanishing
equal amplitude of the scalar and electromagnetic fields $\varepsilon = E_0 \neq 0$. Indeed
in this case equations~(\ref{omega-mix-eq}) and~(\ref{m-mix-eq}) give
$m^2 = 2\omega^2 = \bar{\kappa}^2 (\rho + p + \varepsilon^2)$.  \label{foot-5} }

Finally, from equation~(\ref{m-mix-eq}) one has that
the combination of scalar plus electromagnetic field with a perfect fluid
gives rise to ST-homogeneous G\"odel-type solutions in
hyperbolic class ($m^2>0$)  when  $E_0^2 < \rho + p + 2\varepsilon^2$,
in the linear family  ($m^2=0$) for $E_0^2 = \rho + p + 2\varepsilon^2$,
and also in the trigonometric class ($m^2<0$) when
$E_0^2 > \rho + p + 2\varepsilon^2$.
Moreover, from equation~(\ref{m-mix-eq}) we also have that these three
families of ST-homogeneous G\"odel-type solutions can also be generated
with a simple combination of scalar and electromagnetic fields (thus for $\rho =p=0$),
depending on the relative values of the amplitudes $\varepsilon^2$ and $E_0^2$,
which give the sign of the term  $2\varepsilon^2 - E_0^2$ in equation~(\ref{m-mix-eq}).

\section{Concluding remarks}

Despite the great success of the general relativity theory, a good deal of
effort has recently gone into the study of the so-called modified 
gravity theories. In the cosmological modeling framework, this is
motivated by the fact that these theories provide an alternative way to
explain the current accelerating expansion of the Universe with no need to
invoke the dark energy matter component.
The hybrid metric-Palatini $f(\mathcal{R})$ gravity is a recently devised approach to
such modified theories, in which it is added to the ordinary Ricci scalar $R$, in the
Einstein-Hilbert Lagrangian, a function, $f(\mathcal{R})$, of Palatini curvature
scalar $\mathcal{R}$, which is constructed from the independent connection
$\Gamma_{\mu\nu}^{\rho}$.

In general relativity, the bare existence of the G\"odel solution to Einstein's
equations, for a physically well-motivated perfect-fluid source,  makes it apparent
that this theory permits solutions with violation of causality on nonlocal scale, regardless
of its local Lorentzian character that ensures the local validation of the causality
principle.
In the context of the hybrid metric-Palatini $f(\mathcal{R})$ gravity theories
the space-time manifolds are also assumed to be locally Lorentzian. Hence
the chronology and causality structure of special relativity are inherited
locally.  The nonlocal question, however, is left open,  %
and violation of causality can in principle arise.

Since  homogeneous G\"odel-type geometries necessarily lead to the
existence of closed timelike circles (Section~\ref{CTLC}), which is an unequivocal
manifestation of violation of causality, a natural way to tackle this question
is by investigating whether the hybrid metric-Palatini $f(\mathcal{R})$ gravity
theories permit G\"odel-type solutions to their field equations.
Furthermore, if gravity is to be governed by a $f(\mathcal{R})$  there are
a number of issues ought to be reexamined in its context, including the
question as to whether these theories admit G\"odel-type solutions,
or would remedy the violation of causality problem by ruling out this
type of solutions, which are permitted in general relativity.

In this article, to proceed further with the investigations on the potentialities,
difficulties and limitations of $f(\mathcal{R})$,  we have examined whether
$f(\mathcal{R})$ gravity theories admit homogeneous G\"odel-type solutions for
physically well-motivated matter sources.
To this end, we have first examined the problem of finding out ST-homogeneous
solutions in these hybrid gravity theories  whose trace $T$ (invariant) of
the energy-momentum tensor is constant. We have shown that under this assumption
the problem of finding out solutions in the hybrid metric-Palatini $f(\mathcal{R})$
gravity reduces  to the problem of determining ST-homogeneous solutions
of Einstein's field equations with a cosmological constant.
Employing this far-reaching simplifying result, we first have found
a general G\"odel-type solution for a source that is a combination of scalar
and electromagnetic fields with a perfect fluid.
In this general G\"odel-type solution solution the essential parameter $m^2$
can be $m^{2} > 0$ (hyperbolic family), $m=0$ (linear class) or $m^{2} < 0$
(trigonometric family) depending on the values of the amplitudes
$\varepsilon$ (scalar field) and $E_0$ (electromagnetic field), and the
density $\rho$ and pressure $p$ of the perfect fluid.
This general homogeneous G\"odel-type solution also contains previously known
solutions of the literature as special cases.
There emerges from one of the particular solution of the hyperbolic family
that every perfect-fluid G\"{o}del-type solution of any  $f(\mathcal{R})$
gravity with density $\rho$ and pressure $p$ and satisfying the weak energy
conditions $\rho >0 \; \mbox{and}\; \rho+p \geq 0$ is necessarily isometric
to the G\"odel geometry.
This extends to the context of  $f(\mathcal{R})$  gravity a
theorem established in the context of the general relativity,
which states that G\"odel solution is the sole perfect fluid
solution of Einstein's equations~\cite{BampiZordan78}.

Whether or not the physical laws permit the existence of stable time machines
in the form of closed timelike curves is a research field in general relativity
and other gravity theories.
Violation of causality on the other hand raises intriguing logical paradoxes,
and is generally seen as undesirable feature in physics.  
The two most known remedies to these paradoxes are Novikov's self-consistency
principle~\cite{Novikov,Carlini_etal-1995,Carlini_etal-1996},
which was designed to reconcile the logical inconsistences by demanding that
the only admissible local solutions are those which are globally self-consistent,%
\footnote{Clearly this principle is ultimately an ad hoc global topological constraint
on admissible solutions of gravity theories, thus beyond the standard scope of the
local formulation of the gravity theories.}
and Hawking's chronology protection conjecture\cite{Hawking1992}, which suggests
that even though closed timelike curves are classically possible to be produced,
quantum effects are likely to prevent such time travel. In this way, the
laws of quantum physics  would prevent closed timelike curves from appearing.%
\footnote{For a good pedagogical overview with a fair list of references on
the chronology protection conjecture and Novikov's self-consistency
principle see  Visser~\cite{M_Visser_2003}, and the recent review article
on closed timelike curves and violation of causaltiy by Lobo~\cite{Francisco-Lobo-2012}}
In this regard, Hawking and Penrose have also pointed out that severe causality assumptions
could risk 'ruling out something that gravity is trying to tell us'~\cite{Hawking-Penrose-1996},
thus, discouraging further investigations. 
The possible existence of closed timelike curves is also particularly interesting in the
quantum realm, where, for example, the quantum systems traversing these curves have
been studied~\cite{Deutsch-1991} and experimental simulation of closed timelike
curves have been undertaken~\cite{Ringbauer-etal-2014}.

To conclude, we emphasize that  the bare existence of the ST-homogeneous G\"odel-type
solutions that we have found makes apparent that the hybrid metric-Palatini $f(\mathcal{R})$
gravity does not remedy at a classical level the causal pathology
in the form of closed timelike curves that arises in the context of
general relativity. We are not aware of a quantum gravity theory
following the hybrid metric-Palatini structure, though.


\begin{acknowledgements}
M.J. Rebou\c{c}as acknowledges the support of FAPERJ under a CNE E-26/202.864/2017
grant, and thanks CNPq for the grant under which this work was carried out.
J. Santos acknowledges support of Programa de P\'{o}s-Gradua\c{c}\~{a}o em
F\'{i}sica - CCET/UFRN.
\end{acknowledgements}



\end{document}